\documentclass[runningheads,anonymous]{llncs}
\usepackage{amsmath,graphicx,multicol,multirow,booktabs,amssymb,bbding}
\usepackage{graphicx}
\usepackage{algorithm}
\usepackage{algorithmic}
\usepackage[marginal]{footmisc}

\begin{document}
\title{An Optimization-based Baseline for Rigid
2D/3D Registration Applied to Spine Surgical
Navigation Using CMA-ES}
\titlerunning{An Baseline for Rigid 2D/3D Registration Using CMA-ES}
\author{Minheng Chen\inst{1}$^\#$ \and
Tonglong Li\inst{1,2}$^\#$ \and
Zhirun Zhang\inst{1}$^\#$ \and
Youyong Kong\inst{1,2,3(}\Envelope\inst{)}
}
\authorrunning{M. Chen, T. Li and Z. Zhang et al.}
\institute{School of Computer Science and Engineering, Southeast University, Nanjing 210096, China \and
Jiangsu Provincial Joint International Research Laboratory of Medical Information Processing,  Southeast University, Nanjing 210096, China \and
Key Laboratory of New Generation Artificial Intelligence Technology and  Its Interdisciplinary Applications (Southeast University), Ministry of Education, Nanjing 210000, China\\
\email{ kongyouyong@seu.edu.cn}\\
}
% %
\maketitle              % typeset the header of the contribution
\footnote{$^\#$The three authors contributed equally to this work.}
\begin{abstract}
A robust and efficient optimization-based 2D/3D registration framework is crucial for the navigation system of orthopedic surgical robots. It can provide precise position information of surgical instruments and implants during surgery.
While artificial intelligence technology has advanced rapidly in recent years, traditional optimization-based registration methods remain indispensable in the field of 2D/3D registration.
% A refinement step employing classical optimization-based 2D/3D registration methods combined with deep learning-based techniques can deliver the necessary level of accuracy.
The exceptional precision of this method enables it to be considered as a post-processing step of the learning-based methods, thereby offering a reliable assurance for registration.
In this paper, we present a coarse-to-fine registration framework based on the CMA-ES algorithm.
We conducted intensive testing of our method using data from different parts of the spine. The results shows the effectiveness of the proposed framework on real orthopedic spine surgery clinical data. 
\textbf{This work can be viewed as an additional extension that complements the optimization-based methods employed in  our previous studies.}
\end{abstract}
\section{Introduction}
Automatic X-ray to CT registration is a process that aims to align intra-operative X-ray images with corresponding pre-operative CT scans. 
It involves finding the spatial correspondence between these two modalities, enabling accurate integration and analysis of information from both imaging techniques. The challenges in automatic X-ray to CT registration arise due to differences in image acquisition protocols, patient positioning, and image artifacts. Additionally, anatomical deformations caused by patient movement or pathological changes present further complexities. 
And it has shown promising results in various clinical applications, including orthopedics, interventional radiology and minimally invasive surgical robot navigation. It allows clinicians to effectively fuse the information from X-ray and CT modalities, providing a comprehensive understanding of a patient's condition and facilitating more precise and targeted medical interventions~\cite{unberath2021impact}.

Recent progress in machine learning has had a significant impact on 2D/3D registration, revolutionizing the field and improving the accuracy and efficiency of the registration process~\cite{fu2020deep}. 
Researchers have started exploring the use of neural networks as a substitute for traditional similarity measures~\cite{neumann2020deep}, treating registration as a Markov decision process~\cite{miao2018dilated}, and employing differentiable projection operators to directly implement an end-to-end registration framework~\cite{gao2023fully,gopalakrishnan2022fast}. 

Some existing works~\cite{gopalakrishnan2023intraoperative,zhang2023patient} get rid of the problem of lack of real data by adopting self-supervised training strategies.
However, in the existing literature, Most learning-based registration methods still require the use of optimization-based methods as a post-processing step to fine-tune the results. For example, \cite{chen2024fully,gao2023fully} use neural networks to obtain an approximately convex mapping, which can increase the capture range of registration. But this network similarity function is overly smooth, thereby leading to premature convergence when the pose closely approximates the ground truth.  In order to ensure the accuracy of registration, a benchmark based on covariance adaptive evolution strategy (CMA-ES)~\cite{hansen2003reducingCMAES} is adopted for refinement. 
Gao et al.~\cite{gao2020generalizing}, Gopalakrishnan et al.~\cite{gopalakrishnan2022fast} and  Zhang et al.~\cite{zhang2023patient} all proposed differentiable renderer and employed the gradient descent optimization method to refine the pose using this module. 
This implies that an efficient and robust optimization-based registration method is still beneficial to the existing registration framework.

In this work, we proposed a coarse-to-fine benchmark for 2D/3D registration.  The framework uses CMA-ES as the optimizer and is divided into two resolutions for pose estimation. 
We validate our proposed framework on vertebral data, demonstrating its ability to achieve high registration accuracy. 
% In addition, we combine our method with some existing learning-based methods and report the performance of the proposed method when using learning-based methods as pose initializer.
Our paper is organized as follows: Sect.~\ref{section::related} provides an overview of related work, Sect.~\ref{section::method} describes the proposed method. And in Sect.~\ref{section::experiment}, we present our experimental setup, datasets, quantitative and qualitative results, and analysis.
\textbf{This work can be seen as a supplementary note on the optimization-based methods we used in~\cite{chen2024fully,chen2023embedded}.}
\section{Related Work}
\label{section::related}
\subsection{Intensity-Based 2D/3D Registration}
 In intensity-based methods, a simulated X-ray image, referred to as Digitally Reconstructed Radiograph (DRR), is derived from the 3-D X-ray attenuation map by simulating the attenuation of virtual X-rays.
 An optimizer is employed to maximize an intensity-based similarity measure, such as normalized cross-correlation (NCC) and mutual information,  between the DRR and X-ray images. Common mathematical optimization methods for 2D/3D registration include Powell-Brent~\cite{powell1964efficient}, Nelder-Mead, nonlinear conjugate gradient, gradient descent, evolutionary strategy, etc~\cite{van2011evaluation}.
 It is widely recognized that intensity-based methods~\cite{grupp2018patch} can achieve high registration accuracy. However, these methods also have two significant drawbacks: long computation time and limited capture range. In recent years, many literatures  have tried to use neural networks as pose initialization for intensity-based methods~\cite{gao2023fully,gopalakrishnan2023intraoperative,zhang2023patient}. Learning-based methods can often initialize poses near the ground truth, which makes up for the shortcomings of the smaller capture range of intensity-based methods.
\subsection{Feature-Based 2D/3D Registration}
Feature-based methods calculate similarity measures efficiently from geometric features extracted from the images, e.g., corners, lines and segmentations, and therefore have a higher computational efficiency than intensity-based methods. One potential drawback of feature-based methods is that they heavily rely on accurate detection of geometric features, which in itself can be a challenging task.
Errors from the feature detection step are inevitably propagated into the registration result, making feature-based methods in general less accurate. Errors from the feature detection step inevitably propagate into the registration result, generally compromising the accuracy of feature-based methods.
\section{Methodology}
\label{section::method}
Our registration framework is divided into two stages: coarse registration and fine registration, both of which use CMA-ES as the optimizer. Coarse registration is performed on 4$\times$downsampled images (256$\times$256), and fine registration is performed on the original full-resolution (1024$\times$1024). In the coarse registration stage, we use multi-scale normalized cross-correlation (mNCC) as the similarity function, while the fine registration method uses gradient correlation (GC). 
In the following part of this section, we will first introduce the problem formulation of this task and make a brief introduction on the adopted optimizer, CMA-ES. We will also discuss the similarity functions we used for the proposed framework.
\subsection{Problem Formulation}
The problem of rigid 2D/3D registration can be formulated as follows: Given a fixed 2D X-ray image $ I \in \mathbb{R}^{H\times W} \longrightarrow \mathbb{R}$  and a moving 3D volume $V\in\mathbb{R}^{H\times W\times D}\longrightarrow \mathbb{R}$ as input.  We aim to seek an unknown camera pose $\theta \in \textbf{SE}(3)$ such that the image projected from $V$ is as similar as possible to the acquired image $I$. It is important to note that in this study, the three-dimensional volume $V$ used is a segmentation of vertebra, as bone is a rigid object with higher attenuation than soft tissue, making it more suitable for feature extraction.
% This can be expressed as: $\textbf{T}\longrightarrow \mathop{\arg\min}\limits_{\textbf{T}}(\mathcal{P}(\textbf{T};V),I)$
% a mapping function $\mathcal{F}$ to retrieve the pose parameter $\theta \in \mathrm{SE(3)}$ : 
\subsection{Optimizer}
CMA-ES is an evolutionary strategy designed for continuous optimization problems. It is a probabilistic-based optimization method that simulates natural selection and genetic mechanisms of biological evolution to optimize parameters.
The core idea of CMA-ES is to search for the optimal solution by gradually adjusting the probability distribution of the parameters. In each generation, CMA-ES generates a set of candidate solutions based on the current probability distribution and updates the distribution according to their performance. By iteratively optimizing the probability distribution, CMA-ES can effectively explore the search space and find better solutions.
CMA-ES performs well in handling high-dimensional, non-convex optimization problems and exhibits robustness and convergence properties compared to other optimization algorithms. A public
implementation of CMA-ES can be found here$^1$\footnote{$^1$https://github.com/CyberAgentAILab/cmaes}

In our framework, if the current similarity function is below a predefined threshold, the registration is considered to have converged. We also set up an additional early stopping strategy. If the minimum value of the similarity loss hasn't been updated after 100 generations of sampling, the registration process will be terminated immediately.
\subsection{Similarity Functions}
\subsubsection{Multi-scale normalized cross-correlation.} 
Normalized cross correlation (N-CC) is a widely-used metric for image similarity measurement. 
It can be expressed as follows:
\begin{equation}\label{eq1}
NCC(I_1,I_2)=\sum_{i=0}^m\sum_{j=0}^n\frac{(I_1(i,j)-\bar{I_1})(I_2(i,j)-\bar{I_2})}{\sigma_{I_1}\sigma_{I_2}}
\end{equation}
where $I_1$ and $I_2$ are two images of size $m\times n$. $\sigma_{I_1}$, $\sigma_{I_2}$ represents the standard deviations of $I_1$ and $I_2$, $\bar{I_1}$, $\bar{I_2}$ denote the mean of the image intensities.The NCC calculated directly on the entire image is commonly known as global NCC~\cite{van2011evaluation}.
Patch-based NCC~\cite{grupp2018patch} is also a common similarity function, which is also called local NCC. In this work, we only consider square shaped patches, defined by the patch center($p_x$,$p_y$) and a radius, $r$. And it can be formulated as:
\begin{equation}\label{eq2}
LNCC(I_1,I_2,p_x,p_y,r)=\sum_{i=p_x-r}^{p_x+r}\sum_{j=p_y-r}^{p_y+r}\frac{(I_1(i,j)-\bar{I_1})(I_2(i,j)-\bar{I_2})}{\sigma_{I_1}\sigma_{I_2}(2r+1)^2}
\end{equation}
$\sigma_{I_1}$, $\sigma_{I_2}$ represents the standard deviations of the corresponding patches in $I_1$ and $I_2$.
Multi-scale NCC is a hybrid metric that combines the two aforementioned metrics. Assuming that the image is divided into K patches, the multi-scale NCC can be mathematically expressed as:
\begin{equation}\label{eq3}
mNCC(I_1,I_2)=NCC(I_1,I_2)+\lambda\sum_{(p_i,p_j)\in\Omega_K}LNCC(I_1,I_2,p_i,p_j,r)
\end{equation}
$\lambda$ is a hyperparameter and in this work we set it to 1. As for patch radius $r$, we set it to 6 during experiment . 
Compared with global NCC, multi-scale NCC is more sensitive to texture details. And it is more stable than local NCC and less likely to fall into local minima. A public implementation of mNCC can be found here$^2$\footnote{$^2$ https://github.com/eigenvivek/DiffDRR}.

In addition, we also considered using the intensity variance weighting method to give weight to each patch like some previous works~\cite{grupp2018patch,knaan2003effective}. However, we discovered that this approach led to an unstable registration effect, especially noticeable in images with high noise levels or complicated anatomical regions like the cervical vertebrae.
\subsubsection{Gradient correlation.} Gradient-based measures initially transform $I_1$ and $I_2$ by differentiation. We utilize horizontal and vertical Sobel templates to generate gradient images, $dI_1/di$ and $dI_1/dj $,  representing the derivative of fluoroscopy intensity along the two orthogonal axes of the image. Subsequently, normalized cross correlation is then calculated between $dI_1/di$ and $dI_2/di$ and between $dI_1/dj$ and $dI_2/dj$. The final value of this measure is the average of these normalized cross correlations.
\begin{equation}\label{eq4}
GC(I_1,I_2)=\frac{NCC(dI_1/di,dI_2/di)+ NCC(dI_1/dj,dI_2/dj)}{2}
\end{equation}
GC exhibits a sharp peak at the ground truth camera pose, but its landscape contains numerous local minima. On the other hand, mNCC is substantially smoother but has less defined peaks. As a result, we adopt mNCC as the similarity function during the coarse registration stage and subsequently replace it with GC during the fine registration stage.
\section{Experiments}
\label{section::experiment}
\subsection{Dataset and Experiment Environment}
\subsubsection{Dataset.}
We employed fifteen spine CT scans to evaluate the performance of the proposed method, comprising five cervical spine scans, five thoracic spine scans, and five lumbar spine scans.
Each scan has a corresponding X-ray with Postero-Anterior (PA) views.
For coarse registration, the size of the x-ray used is  256 $\times$ 256 ($4\times $ downsampled), and fine registration uses the original image resolution.
\subsubsection{Image pre-processing.}
For each X-ray, the ground truth extrinsic matrix is provided and a logarithmic transformation following Beer-Lambert's law is applied to invert the intensity of the image. 
The spines were segmented using an automatic method in~\cite{zhang2024spineclue}.
To ensure consistent and standardized analysis, we employed a resampling technique on the CT scans, resulting in an isotropic spacing of 1.0 mm. 
Additionally, we applied a cropping or padding process along each dimension, yielding volumes of size $256\times256\times256$, with the spine ROI approximately positioned at the center of the volume. 

% \subsubsection{Implementation Details.}
\subsubsection{Experiment settings.}
The camera intrinsic parameters used in the experiments simulate a Perlove PLX118F mobile C-arm imaging device which generates the X-ray images in this work.
The device has an isotropic pixel spacing of 0.19959 mm/pixel, a source-to-detector distance of 1011.7 mm, and a detector dimension of 1024 $\times$ 1024 pixels.
For each subject, twenty registrations were performed using initial poses sampled from normal distributions of $N (0, 20)$ for rotations in degrees and $N (0, 30)$ for translations in millimeters.
% In addition, we also conducted orthogonal-view registration experiments on lumbar spine data. 
% The transformation parameters from the AP view to the lateral view of each pair of X-rays are known.
% And we set the weight ratio of the similarity function from the front and lateral images to 7:3.
\subsection{Evaluation Metrics}
Following the standardized evaluation methods in 2D/3D registration~\cite{van2005standardized}, we report mean target registration error (mTRE) in 50th, 75th, and 95th percentiles (in millimeters). 
mTRE is defined as the average 3D distance between the projections obtained from the ground truth camera poses and the estimated camera poses. Suppose we have a three-dimensional point set $\Pi$consisting of $L$ anatomical landmarks, mTRE can be represented as:
\begin{equation}\label{eq5}
mTRE(\theta,\hat{\theta})=\frac{1}{L}\sum_{v_i\in\Pi}^L\Vert\theta\circ v_i-\hat{\theta}\circ v_i\Vert_2
\end{equation}
% We define a condition for successful registration as mTRE less than 10mm, and report the success rate of registration.
We also evaluate the errors in rotation and translation between estimated $\theta$ and ground truth $\hat{\theta}$ respectively.
\subsection{Results}
The numerical results of the registration pose error are shown in Table.~\ref{result}. Because our experiments were initiated with rather substantial offsets, the mean errors were significantly skewed by the presence of large-scale outliers that do not truly reflect the actual distribution. The initial mTRE is $118.74\pm54.36$, $98.59\pm41.35$, and $75.40\pm29.89$ at the 95th, 75th, and 50th percentiles respectively. Because the sizes of different anatomical parts of the spine are different, the mTRE obtained from the experimental results on cervical, thoracic and lumbar spine data varies greatly. It will be more intuitive to directly compare the errors of each component in rotation and translation. It is worth noting that although the mean values of errors in some directions of rotation and translation became larger after registration in our experiments, this was actually affected by outliers. 
Taking the total errors in the three directions of rotation (rx, ry, rz) as an example: their initial errors are $4.02\pm3.16$, $3.79\pm3.13$ and $3.76\pm2.93$, while the errors after registration are $4.21\pm7.46$, $6.06\pm7.86$ and $2.92\pm5.11$. 
And the medians of the initial errors are 3.10, 2.90, and 3.16, while the medians of the registration results are noticeably smaller, measuring 0.52, 1.32, and 0.24 respectively.
\begin{table}[h!]
%\footnotesize
\begin{center}
	\label{table1}
 
	\caption{2D/3D registration performance  on cervical, lumbar and thoracic spine data. This evaluation includes measurement of the errors in rotation and translation, the mean Target Registration Error (mTRE) at the 50th, 75th, and 95th percentiles.}
 \setlength{\tabcolsep}{0.6mm}{
	\begin{tabular}{c|l|c|c|c|c|ccc|c|ccc} 
  \toprule[1.5pt]
 \multirow{2}{*}{} & \multirow{2}{*}{Subject}
  & \multicolumn{3}{|c|}{mTRE(mm)$\downarrow$}& \multicolumn{4}{l}{Rotation Error($^\circ$)$\downarrow$}& \multicolumn{4}{|l}{Translation Error(mm)$\downarrow$}
		\\
 \cline{3-13}    &  & 95th & 75th & 50th&rotate. &rx
  &ry &rz&trans.&tx&ty&tz\\
  \hline
  \hline
  \multirow{5}{*}{\textbf{Cervical}}& $\#$1 & 136.8 &101.1 &  74.2& 34.9 & 21.9&6.7& 6.4&13.5&2.1 & 4.8& 6.7\\
  & $\#$2 & 319.3&266.4 &  213.6 &27.7& 6.3 &13.8 & 7.5& 47.5&7.0&20.1&20.3 \\

& $\#$3 &369.0 &332.0 &292.0 &34.1&9.7&12.3&12.1 & 31.4&5.0&8.8&17.7\\
 & $\#$4 &367.5 &335.0 &291.8&38.1&8.9&19.6& 9.6&58.3&16.8&10.6&30.9\\

& $\#$5 &306.5 & 275.5 &240.8 &27.3&6.0&17.5&3.8  &46.2&3.6&22.0&20.5\\
& $\#$1-5 &296.5 &251.9&191.6 &32.4 &10.6&14.0&7.9 & 39.4&7.4&12.7&19.2\\
\hline
\hline
\multirow{5}{*}{\textbf{Thoracic}}&$\#$6 &20.5 & 20.4 &20.2& 1.4& 0.3&0.9 &0.2&7.7&0.2&1.7&5.8\\

& $\#$7 & 25.6& 25.1& 24.0&1.4& 0.3& 1.1& 0.1&6.5& 0.3&1.0&5.3\\
& $\#$8 &28.0 & 20.4 &20.4& 3.2& 0.2&2.5&0.6 &9.4&0.2&3.1&6.1\\

& $\#$9 &46.8&31.3 &30.6&6.3&2.3&3.1&0.9&13.4 &0.9& 6.2&6.3\\
& $\#$10 &10.2 &10.2 &10.1 & 2.5& 1.6&0.8&0.2&4.9&0.4&1.6&2.9\\
% \cline{2-13}
& $\#$6-10 &23.4&19.1&16.4& 3.0& 0.9&1.7& 0.4 & 8.4&0.4&2.7&5.3\\
\hline
\hline
\multirow{5}{*}{\textbf{Lumbar}}&$\#$11 &36.4 &35.4&35.2 &1.5&  0.2 &1.3&0.1&14.6&0.2&2.1& 12.3\\

& $\#$12 & 13.5& 13.4& 13.0&2.3&0.9&1.1&0.2&2.1&0.4 &1.0&0.7\\
& $\#$13 &53.0&25.2&16.1&  6.0& 2.7&2.9& 0.5&15.1&1.9&9.3&3.8\\

& $\#$14 &135.5&128.7&113.2& 10.1&1.7&6.8&1.5 &29.6&1.3&9.6&18.7\\
& $\#$15 &12.9&12.9 &12.9& 1.2&0.3&0.7&0.2&3.9&0.5&1.1&2.3\\
% \cline{2-13}
& $\#$11-15 &47.0 &22.3&14.3 & 4.2 &1.2&2.6&0.5&13.1& 0.9&4.7&7.6\\
\hline
\hline
\multirow{2}{*}{\textbf{Total}}& Initial &118.7&98.6&75.4& 11.6&4.0&3.8&3.8 &23.8&8.1&7.9&7.8\\
& Result &114.6&52.2 &19.8& 13.2&4.2&6.1&2.9&20.3&2.9&6.7&10.7
\\
\bottomrule[1.5pt]
\end{tabular}}
\vspace{-1.0cm}
\label{result}	
\end{center}
\end{table}

The performance of our framework on lumbar and thoracic spine data is very convincing, which hints at the feasibility of this framework in clinical application.  But we also noticed its unsatisfactory performance in cervical spine data. 
We believe that this is mainly due to two reasons: 1)The cervical spine area contains a greater number of joints and bone structures, and has a wider range of motion compared to other spinal regions. As a result, cervical spine images may exhibit a more intricate shape and structure, and the registration process must consider a broader range of variations and uncertainties. 
In contrast, the lumbar and thoracic regions are comparatively larger and have relatively simple structures, so the registration process may be easier.
2)The patient's head direction was not entirely consistent during preoperative and intraoperative imaging, leading to deformations in certain parts of the cervical spine that exceeded the 6 DoF rigid body transformation limit. 
We can mitigate the impact of jaws with significant shape differences in the image by partitioning the region of interest.
However, in such cases, adopting regularization of rigid bodies for cervical spine registration may result in a higher likelihood of falling into local minima.
\section{Conclusion}
Single-view 2D/3D registration inherently has ambiguities in translation (tz) in the depth direction and out-of-plane rotation (rx and ry).
Our method cannot avoid this defect, but other research results~\cite{gao2023fully,gao2020generalizing} show that combining optimization-based methods with learning-based methods can effectively alleviate this problem. 
Our experiments in previous work~\cite{chen2024fully} also substantiate this conclusion. 
In addition, we aspire to develop a more elegant and rational solution to address this problem in future endeavors.

In this paper, we propose a multi-resolution 2D/3D registration algorithm using the CMA-ES algorithm. We verified the effectiveness of this framework using paired CT and X-ray images from three different anatomical sites (lumbar, thoracic, and cervical vertebrae) in the context of spinal surgical navigation.
Our experimental results have yielded highly competitive outcomes. We aim for this method to serve as a benchmark, coupled with learning-based registration methods, and to potentially be implemented in clinical surgical settings in the future.

\section{acknowledgement}
This work was supported in part by Bond-Star Medical Technology Co., Ltd..  
We thank Sheng Zhang,  Junxian Wu and Ziyue Zhang for their constructive suggestions at several stages of the project.

\bibliographystyle{splncs04}
\bibliography{mybib}
\end{document}